 \documentclass[preprint2]{aastex}

\usepackage{graphicx}
\usepackage{amsmath}

\begin{document}

\title{From planetesimals to dust: Low gravity experiments on recycling solids at the inner edges of protoplanetary disks}

\author{Caroline de Beule, Thorben Kelling, Gerhard Wurm, Jens Teiser and Tim Jankowski}
\affil{Universit\"at Duisburg-Essen, Fakult\"at f\"ur Physik, Lotharstra\ss e 1,D-47057 Duisburg, Germany}
\email{caroline.de-beule@uni-due.de}

\begin{abstract}

Transporting solids of different sizes is an essential process in the evolution of protoplanetary disks and planet formation. Large solids are supposed to drift inward; high-temperature minerals found in comets are assumed to have been transported outward. From low-gravity experiments on parabolic flights we studied the light-induced erosion of dusty bodies caused by a solid-state greenhouse effect and photophoresis within a dust bed's upper layers. Tje gravity levels studied were 0.16$g$, 0.38$g$, 1$g$, and 1.7$g$. The light flux during the experiments was 12 $\pm$ 2 kW m$^{-2}$ and the ambient pressure was 6 $\pm$ 0.9 mbar. Light-induced erosion is strongly gravity dependent, which is in agreement with a developed model. In particular for small dusty bodies ((sub)-planetesimals), efficient erosion is possible at the optically thin inner edges of protoplanetary disks. Light-induced erosion prevents significant parts of a larger body from moving too close to the host star and be being subsequently accreted. The small dust produced continues to be subject to photophoresis and is partially transported upward and outward over the surface of the disk; the resulting small dust particles observed over the disk's lifetime. The fraction of  eroded dust participates in subsequent cycles of growth during planetesimal formation. Another fraction of dust might be collected by a body of planetary size if this body is already present close to the disk edge. Either way, light induced erosion is an efficient recycling process in protoplanetary disks. 

\end{abstract}

\keywords{methods: laboratory-, planets and satellites: formation - protoplanetary disks}

\section{Introduction}
The formation of planets in protoplanetary disks is strongly tied to the existence of dust which makes up about 1\% of the total mass of the disk \citep{natta2007}. The basic collisional growth model assumes that, as a first step toward planet formation, kilometer-size planetesimals are formed by sticking collisions of dust particles which are only held together by surface forces \citep{blum2008, weidenschilling1993,windmark2012}. There are some issues in planetesimal formation which are not yet understood. Somewhat problematic is inward drift especially of meter-size bodies. In a standard disk the pressure is supposed to decrease with distance from the star \citep{hayashi1985}. The gas is therefore supported by a pressure gradient and rotates with sub-Keplerian velocity. Small solids couple to the gas and thus also rotate more slowly than Keplerian. However, solid bodies are not supported by the pressure gradient. This leads to an inward drift of the solid bodies which -- for a meter size body in a minimum mass nebula -- can be as large as 1 AU in 100  years \citep{weidenschilling1977}. Such an inward mass transport of solids leads to a redistribution of matter: the outer part of the disk becomes depleted in solids, while the inner part becomes enriched. The physics of the inner 1 AU of protoplanetary disks strongly depends on the fate of this incoming matter: it might contribute to the local planet formation process and add mass to forming planetary bodies, it might evaporate and be accreted by the star, or it might be recycled in some more complex way. As the physics in the inner part of the disk
is complex the evolution of solid bodies is still an open question. To address this problem, we started to investigate a mechanism of particle recycling which is especially active in this inner region of a protoplanetary disk and, more specifically, close to the inner edge of a disk within the inner 1 AU. There is a growing number of observations which show that, in some phases, protoplanetary disks contain inner gaps which are optically thin but still contain gas, while the outer part is still dense and optically thick \citep{dalessio2005,calvet2002,currie2011,sicilia2008}. In other words, solids at the inner edge of a disk are embedded in a gaseous environment and are illuminated by the star. \\

\citet{wurm2006} first described a mechanism that significantly erodes dusty bodies to small particles under very general conditions, e.g. by mere illumination and at low gas pressure (kW m$^{-2}$ of intensity and mbar gas pressure). \citet{wurm2007} showed that this erosion process also works under the conditions of transitional disks. There will most likely be a small shell at the inner edge of a disk where the conditions for particle erosion are met \citep{wurm2007}. Every dusty body which enters this zone and which is too small to hold its dust by gravity (sub-planetesimal) is subject to erosion by this process. Different scenarios are possible as to how this contributes to the evolution of the disk. The dust might, for example, be transported upward and outward over the disk by photophoresis. Such a model has been suggested by \citet{wurm2009}. At later times or throughout the optical thin inner parts the dust might also be transported by photophoresis in the midplane \citep{wurm2007, takeuchi2003,herrmann2007,moudens2011,mousis2007}. \citet{kelling2011b} showed that the erosion process is a suitable mechanism to explain the existence of small dust particles which are observed in protoplanetary disks over their entire lifetime. The dust might also be recycled locally and added -- probably more efficiently than by the original larger bodies -- to existing large planetesimals or protoplanetary bodies to boost their growth. These aspects are speculative and not the main focus of this paper which reports on experiments that analyze the gravity dependence of the light induced erosion mechanism providing basic
input for future modeling.\\ 

\begin{figure}
    \centering
    \includegraphics[width=\columnwidth]{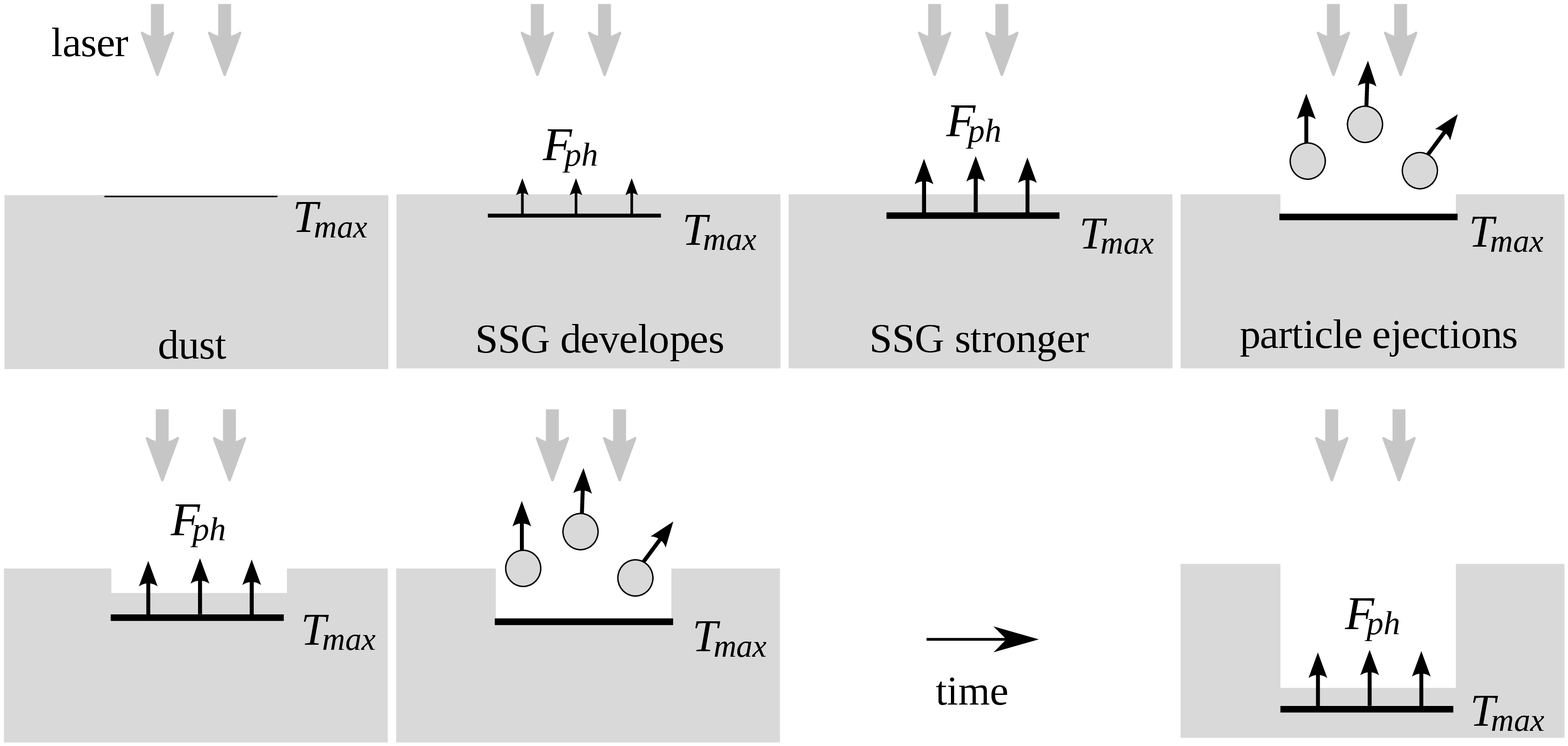}
    \caption{Visible radiation (laser) enters the dust bed and leads to a solid state greenhouse (SSG) effect, with a maximum temperature below the surface (indicated by the black line: larger temperatures are represented by thicker lines). Photophoresis $F_{\text{ph}}$ acts on the surface particles. When this force overcomes gravity and cohesion, the surface particles are lifted. A crater is formed where again temperature gradients develop. If the illuminated spot is limited in size, the developing crater might prevent further release of particle.}
    \label{fig:fig1}
\end{figure}

A sketch of the light induced erosion mechanism is shown in Figure \ref{fig:fig1}: the illumination of a dusty surface first leads to a solid-state greenhouse (SSG) effect. This effect is known to exist for ice and comets \citep{niederdorfer1933, davidsson2002a, kaufmann2002}. It means that visible light enters a porous body and is absorbed over some depth, but as the body is opaque to infrared radiation only the top surface can cool by thermal emission. Therefore, the maximum temperature is below the surface. In recent years it was shown that this is also true for purely dusty surfaces \citep{kocifaj2010,kocifaj2011, kelling2011a,kelling2011b}. If temperature gradients across a particle exist as in the case of an SSG profile, and if the particle is embedded in a gaseous environment, a photophoretic force acts on the particle from the warm to the cold side. Hence, the surface particles of an illuminated dust bed are subject to a lifting force. If this force is stronger than gravity and cohesion, aggregates are ejected. The SSG and photophoresis are the basis for light induced erosion of dusty bodies. The physics behind this has been studied in a number of works \citep{wurm2006,wurm2007,kelling2009,kelling2011a,kelling2011b}.\\

In protoplanetary disks, dust particles on dusty sub-planetesimals are bound by cohesion and gravity. However, the gravitational force is small due to the small mass of the objects. Therefore photophoretic forces acting on the particles have to overcome less gravity and the effect of dust erosion induced by illumination is more intense compared to larger bodies or laboratory experiments. In the first microgravity experiments \citet{wurm2008} showed that the threshold light flux for particle erosion indeed depends on the gravity level. At low gravity cohesion eventually becomes the dominant opposing force. The erosion rate was not measured in the experiments of \citet{wurm2008}: however, this is important in determining how rapidly a body can be eroded or what mass flux it might provide in small particles if it crosses the erosion zone in a protoplanetary disk.\\

To quantify the effect of light induced erosion for application to (sub)-planetesimal recycling we carried out low gravity experiments and determined the gravity dependence of the erosion rate. The experiments were carried out during the first  \textit{Joint European Partial-G Parabolic Flight} campaign. On this campaign, over 3 days, 13 parabolas at 0.16$g$, 12 parabolas at 0.38$g$ and 6 parabolas at 0$g$  ($g$ $=$ 9.81 m s$^{-2}$) were provided. We also developed a preliminary model which agrees well with the experimental results.

\begin{figure*}
    \centering
    \includegraphics[width=\textwidth]{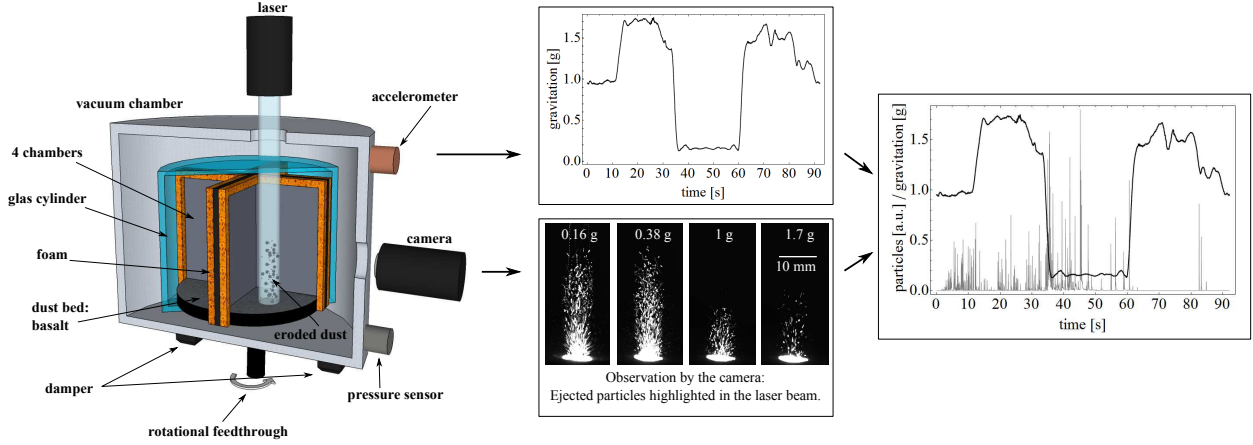}
    \caption{Experimental setup: four chambers filled with dust samples are enclosed by a rotatable glass cylinder in a vacuum chamber. A homogeneous laser spot of 25 $\pm$ 4 mm$^2$ size illuminates the dust bed with an intensity of 12 $\pm$ 2 kW m$^{-2}$. The particles being ejected at different $g$-levels are observed by a side view camera.}
    \label{fig:fig2}
\end{figure*}

\section{Experiment}

A sketch of the experimental setup is shown in Figure \ref{fig:fig2}. Four separate chambers are placed in one vacuum chamber. Each chamber contains a dust sample with a depth of 2 cm covering a total area of 6 cm$^2$. For the campaign reported here the chambers were filled with basalt powder with a broad size distribution between $0$ and $125$ $\mu$m. We consider this dust particle size range suitable to simulate the dust in protoplanetary disks. They include micron-size particles discussed \citet{brearley1999} with reference to the matrix in chondrites (typically a few microns but occasionally up to about 20 $\mu$m \citep{brearley1999}). The constituents in interplanetary dust particles or cometary material are smaller, on the order of 0.1 $\mu$m \citep{wozniakiewicz2012} which might not be adequatley represented in our experiment sample. On the other hand, chondrites largely consist of chondrules of submillimeter size, so the larger grain fraction in our sample might correlate with these dimensions. Certainly, the size dependence of the erosion mechanism should be considered in more detail in the future.\\ 
The dust samples were dried at about 500 K for 24 hr and were stored in a desiccator until they were placed in the vacuum chamber just before the experiments. The ambient pressure of air in the vacuum chamber was then set to $p=6$ $\pm$ 0.9 mbar for all experiments. \\

During each experimental run one sample was illuminated by a diode laser of 2 W optical output at 655 $\pm$ 10 nm. The laser profile was homogenized by coupling the laser to an optical fiber and projecting the fiber outlet onto the sample surface with a size of 25 $\pm$ 4 mm$^2$. The average intensity was 12 kW m$^{-2}$ which varied by 17 \% within the spot. This corresponds to the inner regions of a protoplanetary disk $<$ 0.4 AU where the light flux for a star similar to the sun is  $I > 10$ kW m$^{-2}$.  An accelerometer was used to measure the residual gravity acting on the experiment. The experimental chamber was supported by vibrational attenuators to prevent any influence of airplane vibrations, which might reduce the cohesion between the dust particles. Released aggregates were observed from the side by a camera at 60 images per second.\\ 

\begin{figure}
    \centering
    \includegraphics[width=\columnwidth]{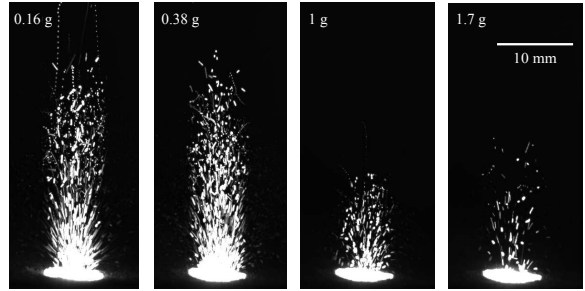}
    \caption{Dust erosion of basalt powder at \mbox{6 mbar} ambient pressure and a light intensity of \mbox{12 kW m$^{-2}$} at four different gravitational accelerations. Each figure is an overlay of 100 images to emphasize the increase of particle release with lower accelerations.}
    \label{fig:fig3}
\end{figure}

\begin{figure}
    \centering
    \includegraphics[width=\columnwidth]{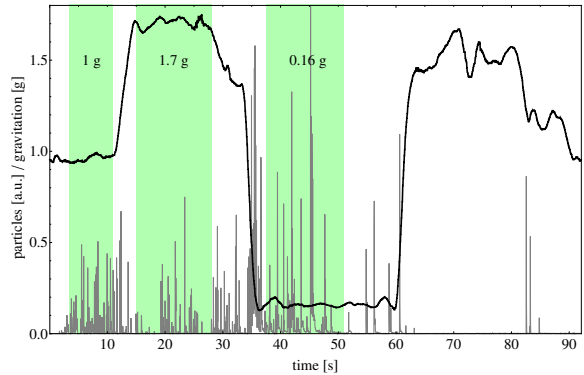}
    \caption{Example of the collected raw data of a Moon parabola. The black line shows the gravitational acceleration in units of \mbox{$g$ = 9.81 m s$^{-2}$}. The gray peaks indicate the eroded mass in arbitrary units. For the analysis only the data in the green boxes were taken where the release of  aggregates is continuous.}
    \label{fig:fig4}
\end{figure}

\begin{figure}
    \centering
    \includegraphics[width=\columnwidth]{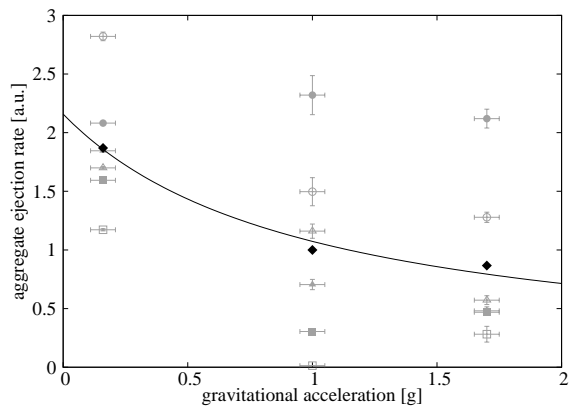}
    \caption{Dataset for the 6 Moon parabolas. The particle erosion rate is set to 1 at 1$g$. The results from the experiments are presented in gray, the average value in black and the black line shows the trend for the averaged data according to the model developed in Section 3.}
    \label{fig:fig5}
\end{figure}

\begin{figure}
    \centering
    \includegraphics[width=\columnwidth]{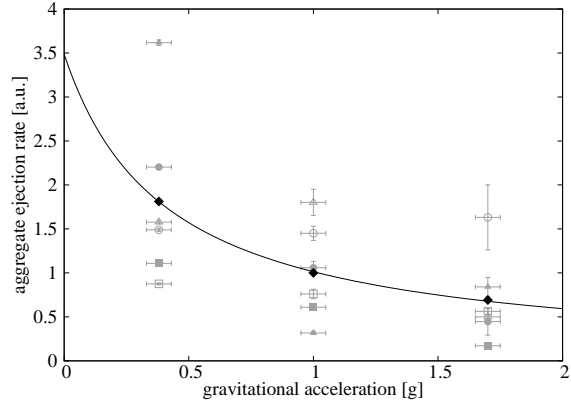}
    \caption{Dataset for the 6 Mars parabolas. The particle erosion rate is set to 1 at 1$g$. The results from the experiments are presented in gray, the average value in black and the black line shows the trend for the averaged data according to the model developed in Section 3.}
    \label{fig:fig6}
\end{figure}

Measurements (images) were taken for 1$g$, the acceleration phase at 1.7$g$ and the low-gravity phase at 0.16$g$ (Moon level) or 0.38$g$ (Mars level). To confine the dust within the individual chambers, they were enclosed in a glass cylinder. This cylinder and the chambers were independently rotatable by a rotational feedthrough from the outside. After each parabola, and hence before every new experiment, the inside of the rotating glass cylinder was cleaned by a small strip of foam. This ensured that the laser and the camera view were not affected by dust.
Rotating the inner, dust containig part of the chambers ensures that for every new experiment a fresh spot of the dust surface was illuminated. This prevented selection effects, e.g., that all ejectable particles (e.g. of a certain size fraction) were already released. After removing the background noise of the camera images the average gray scale value was taken as a relative measurement for the ejected mass or particle number. This assumed that the ejected particle distribution stays constant with time, that averaging is done over a significantly large number of particles and that the image is not saturated with particles. As we provide new spots for each experiment, observe well separated aggregates and take an average over a large number of aggregates in typically 900 images for a given $g$-phase \mbox{(Figure \ref{fig:fig3})}, we consider the gray value as a suitable measure. \mbox{Figure \ref{fig:fig3}} shows an overlay of 100 images of the ejected aggregates at different g-levels. The number of ejected aggregates clearly increases with lower gravitational acceleration. \\
A parabolic flight has the advantage that different gravity levels can be tested with the same dust sample in a short period of time. However, the effects of the transitions between the different $g$-levels have to be filtered out. They induce a higher erosion rate during the transitions from high-$g$ to low-$g$. This is due to the release of gravitational and photophoretic tension from the dust bed, as higher gravity levels require a steeper temperature gradient for erosion. It takes a certain time before the new equilibrium temperature profile is established at which aggregates are ejected regularly. \mbox{Figure \ref{fig:fig4}} shows which data were selected to determine the number of ejected aggregates in the different phases. 
In the 1$g$-phase of every parabola the laser and camera were turned on. First erosions occur after 2-3 s.
During the transition from 1.7$g$ to lower $g$-levels, massive particle releases appear (peaks in Figure \ref{fig:fig3}). These releases are an artifact of the tension release as previously mentioned and are not used to determine the erosion rate.
After the transition particle releases occur continuously at lower $g$-levels for about 10 s. The dust bed surface changes as the laser spot removes particles and creates an erosion crater (see also Figure \ref{fig:fig1}). The presence of a crater rim induces edge effects, which efficiently reduce the erosion rate, especially in the low gravity phase with high erosion rates. Therefore, we do not include the later times for averaging. 
After the data reduction, a measure of the erosion rate is given as the average brightness of all images within a single $g$-phase. As the brightness is an arbitrary measure of the erosion rate,
the different $g$-phases of one parabola are scaled to a value of 1 at 1$g$. \\
The parabolic flight provided results for 12 parabolas: 6 Moon parabolas and 6 Mars parabolas. Figures \ref{fig:fig5} and 6 \ref{fig:fig6} show the data sets for the Moon and Mars parabolas.\\
We further average over all parabolas (6 data points for 0.16$g$ and 0.38$g$ and 12 data points for 1$g$ and 1.7$g$) and obtain the average ejected mass rate over gravity. In Figure \ref{fig:fig8}, all (averaged) experimental data are presented in the context of the model developed below.

\section{The Model}
The continuous aggregate ejections from the dust bed are caused by an inverse temperature  gradient below the dust bed's surface and a photophoretic force $F_{\text{ph}}$ \citep{wurm2006,kelling2011a,kelling2011b}. In detail the idea behind the light induced aggregate ejections is the following: \citet{kocifaj2010,kocifaj2011} showed, that if a dust bed is illuminated from the top, the maximum temperature $T_{\text{max}}$ is up to some 100 $\mu$m below the surface -- this is called SSG. Hence, there is a temperature gradient $\partial T/\partial z$ pointing from $T_{\text{max}}$ toward the cooler surface (inverse temperature gradient; $z$ as depth within the dust bed). Typical thermal conductivities $\kappa_d$ of the dust bed are on the order of $\kappa_d \sim 10^{-2}$ W m$^{-1}$ K$^{-1}$ \citep{krause2011}. The inverse temperature gradient then takes values of $\partial T/\partial z \sim$ 10$^{5}$  K m$^{-1}$ at 10 $\rm $kW m$^{-2}$ illumination \citep{kocifaj2011}. \\
The upper most aggregates have a temperature gradient along their surface. Small particles with a temperature gradient in a gaseous environment are affected by a photophoretic force which can be written for spherical particles as \citep{rohatschek1995}
\begin{eqnarray}
F_{\text{ph}} &=&\frac{2F_{\text{max}} }{\frac{p}{p_{\text{max}}}+ \frac{p_{\text{max}}}{p}}\label{eq:roha1}\\ 
F_{\text{max}}&=& D\sqrt{\frac{\alpha}{2}} a^2\frac{\partial T}{\partial z} \label{eq:roha2}\\
F_{\text{ph}}&=&\omega_1 \frac{\partial T}{\partial z}\label{eq:roha3}
\end{eqnarray} 
where $p$ is gas pressure. $F_{\text{max}}$ and $p_{\text{max}}$ are particle and gas dependent parameters determining the pressure $p_{\text{max}}$ at which the photophoretic force $F_{\text{max}}$ is a maximum. $D$ is a gas dependent parameter, $\alpha=1$ is the accommodation coefficient, $a$ is the particle radius and $\partial T/\partial z$ is the temperature gradient over the particle's surface. Particles in the $\mu$m regime are most strongly affected by photophoresis at mbar pressure.\\

\begin{figure}
    \centering
    \includegraphics[width=\columnwidth]{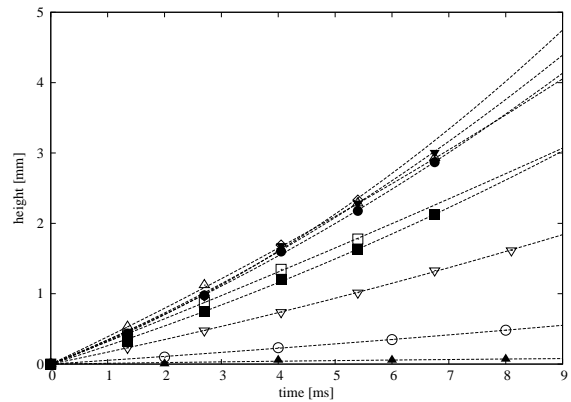}
    \caption{Trajectories of dust  aggregates leaving the surface of an illuminated dust bed (basalt, 0-125 $\mu$m). The experiment was carried out on the ground at 6 mbar ambient pressure and at an intensity of 12 kW m$^{-2}$.}
    \label{fig:fig7}
\end{figure}

\begin{figure}
  \centering
  \includegraphics[width=\columnwidth]{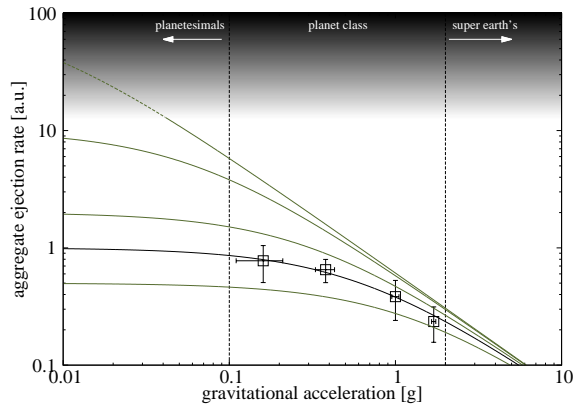}
	\caption{Ejection rate normalized to 1 at 0$g$ for the experimental values using \mbox{Equation \ref{eq:rate}}. The experimental value (squares) for $F_C/\Delta m$ is varied by a factor $f$ with $f=2,1,0.5,0.1,0.01$ (bottom to top). The color gradient indicates that the more aggregates that are ejected, the more light is shielded from reaching the dust bed's surface. From the experiments, we estimate that an increase of the rate up to factor of 10 might be realistic for low cohesive powders.}
	\label{fig:fig8}
\end{figure}

Photophoresis acts on the upper-most aggregates in illuminated dust beds. The threshold for particle ejection can be written as follows \citep{wurm2008}: 
\begin{equation}
F_{\text{ph}_{\text{th}}}=\Delta m \cdot g_{\text{env}}+F_C,\label{eq:threshold}
\end{equation}
with gravity $F_G=\Delta m \cdot g_{\text{env}}$, where $g_{\text{env}}$ is the environmental gravitational acceleration, $\Delta m$ is the ejected mass, and  $F_C$ is the cohesion force. If the photophoretic force $F_{\text{ph}}$ overcomes gravity and cohesion, aggregates are released from the dust bed's surface. The mass ejection rate $N=\Delta m / \Delta t$ with $\Delta t$ as time interval depends on the time scale at which the threshold temperature gradient (see e.g. Equation (\ref{eq:roha2})) within the illuminated dust bed is reset after an ejection. The temperature evolution toward equilibrium within the dust bed follows an exponential dependence \citep{kocifaj2011}. If the threshold gradient to induce particle release is reached after a time span $\Delta t$ much smaller than the exponential timescale (large light flux) the temperature gradient can be approximated by a linear increase with time   
\begin{equation}
{\frac{\partial T}{\partial z}} (t)  = \omega_2 t \label{eq:gradient1}.
\end{equation}
Therefore the threshold temperature gradient after a time span $\Delta t$ can be written as
\begin{equation}
{\frac{\partial T}{\partial z}}_{th} = {\frac{\partial T}{\partial z}} (\Delta t) = \omega_2 \Delta t.\label{eq:gradient2}
\end{equation}
With Equation (\ref{eq:gradient2}) the photophoretic force (Equtation (\ref{eq:roha3})) can be written as
\begin{equation}
F_{\text{ph}_{\text{th}}} = \omega_3 \Delta t
\end{equation}
with $\omega_3 = \omega_1 \omega_2$. Hence, after the time interval $\Delta t$ the threshold photophoretic force ($F_{\text{ph}_{\text{th}}})$, Eq.(\ref{eq:threshold})) is reached
\begin{equation}
F_{\text{ph}_{\text{th}}}=\omega_3 \Delta t = \Delta m \cdot g_{\text{env}}+F_C,
\end{equation}
and aggregates are released in steps always after an average time
\begin{equation}
\Delta t = \frac{ \Delta m \cdot g_{\text{env}}+F_C}{\omega_3} .
\end{equation}
The mass ejection rate therefore is
\begin{equation}
N(g_{env})=\frac{\Delta m}{\Delta t} = \frac{\omega_3}{g_{\text{env}}+\frac{F_C}{\Delta m}}\label{eq:rate}.
\end{equation}
Equation (\ref{eq:rate}) was used to fit the experimental data, and gives \mbox{$\omega_3=1.6$ m s$^{-3}$} and $F_C/\Delta m = 6.1$ $\pm$ 1.2 m s$^{-2}$. $\omega_3$ is scalable arbitrarily as it accounts for the nature of the brightness measurement as the relative measurement for one part. However, $F_C/\Delta m$ contains information on the cohesive force by which an ejected mass fraction is bound to the dust bed. \\
Due to the low resolution of the analyzed images we can not give an accurate assumption for the eroded particle mass $\Delta m$. Hence, we can not estimate the exact value of $F_C$ from our experiments.\\
The acceleration $F_C/\Delta m$ from the model fits the accelerations measured in first follow-up laboratory experiments, shown in Figure \ref{fig:fig7}: basalt powder with a broad size distribution between $0$ and $125$ $\mu$m was placed in a vacuum chamber with 6 $\pm$ 0.9 mbar ambient pressure. It was then illuminated by a laser with 12 $\pm$ 1 kWm$^{-2}$ and a spot size of 5 $\pm$ 0.5 mm radius. The resulting ejection of dust aggregates was measured at a camera with 500 images per second, so the accelerated aggregates could be tracked. We measured accelerations of 14.4 $\pm$ 4.9 m s$^{-2}$, averaged over nine ejected aggregates, which is only twice as high as the acceleration measured on the parabolic flight. This factor two can be explained by the fact that we could not use exactly the same dust sample that we had on board for the parabolic flight. \\
To visualize the influence of the cohesion we plot the ejection rate of Equation (\ref{eq:rate}) divided by the factor $\omega_3 / (F_C / \Delta m)$ for the experimental data (average values of the $g$-phases) which effectively normalizes the experimental data to 1 at 0$g$. 
In Figure \ref{fig:fig8}, the value $F_C / \Delta m$ is then varied by a factor $f$ to show the general dependence on $F_C / \Delta m$.
There is a prominent increase in the ejection rate towards smaller $F_C/\Delta m$ where it becomes more and more dominant at $g_{\text{env}}<1g$ while cohesion is negligible for $g_{\text{env}}>1g$. 

This simple model neglects self-shielding of the dust bed by ejected  aggregates. The more dust that is ejected, the more light is shielded from reaching the surface. Depending on the relation between the ejection rate and the rate at which aggregates are transported away from the ejection location, e.g. by gas drag in protoplanetary disks, the optical thickness of the layer above the dust bed's surface will increase. From the experiments and based on the assumption that the light flux might be reduced to less than $1/e$ of it's intensity, ejection rates that are increased by a factor 10 likely still do not affect the ejection rate. This might set an upper limit to the ejection rate but it is only a rough estimate here which has to be quantified in the future. For planetary bodies the effect might lead to a size sorting over time where large particles fall back down while small grains get entrained into the atmosphere.

\section{Applications and Conclusion}
The effect of light induced dust ejections can be a key process of recycling dust during the formation of planets in protoplanetary disks. Planetesimals or smaller dusty bodies at the inner edge can be partly eroded \citep{wurm2007}.  

\begin{table*}
\begin{tabular}{l c c c c }
\hline\hline
Object&Star's Luminosity ($L_{\text{sun}}$)&Inner Wall (AU)&Erosion Dividing Line (AU)&Outer Wall (AU)\\
\hline
FM 515 & 2.5 &  0.12 &0.63 & 45\\

FM 618 & 2.2 &  0.22 &0.59 & 11\\

LRLL 21 & 3.8 &  0.13 &0.80 & 9\\

LRLL 31 & 5.0 &  0.32 &0.89 & 14\\

LRLL 37 & 1.3 &  0.17 &0.46 & 5\\
\hline\hline
\end{tabular}
\caption{Pre-transitional Disks Modeled by \citet{espaillat2012}. The erosion limit distance is the distance up to which disassembly of larger dusty bodies would be possible.}
\label{tab:table1}
\end{table*}

\begin{figure}
  \centering
  \includegraphics[width=\columnwidth]{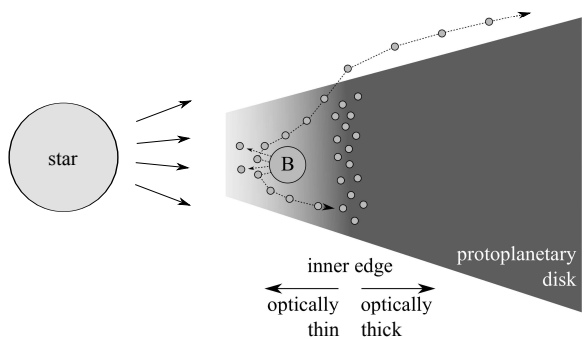}
 \caption{An inward drifting dusty body (B) can be eroded in the optically thin part of a protoplanetary disk by the effect of light induced ejection. The planetesimal loses mass which can be transported outwards again by e.g. photophoresis.}
	\label{fig:fig9}
\end{figure}

The position of the edge in a number of  transition, pre-transition and full disks has been modeled by \citet{espaillat2012}. Here, pre-transition disks relate to a disk with substantial matter within the gap modeled by an optically thin inner wall at the dust sublimation radius and a second outer wall. In Table \ref{tab:table1}, we extract their modeled radii of the disks and the luminosities of the host stars. 
Based on the photophoresis the lift force on particles depends on the environmental parameters light flux $I$, gas pressure $p$ and gas temperature $T$. In the free molecular flow regime (low pressure) Equation \ref{eq:roha1} simplifies and connects these three parameters to
\begin{equation}
F = b\frac{Ip}{T},
\label{eq:lowp1}
\end{equation}
\citep{wurm2007}, with a parameter $b$ dependent on the dust bed. \\
Our experiments show that particle eruptions continuously occur at all gravity levels at the used light flux $I = 12 $ kW m$^{-2}$, pressure $p= 6$ mbar and temperature $T$ of about $300$ K. We did not probe the threshold condition where no dust is lifted but our values are consistent with threshold conditions measured for eruptions at 1$g$ by \citet{wurm2007}. He found $I = 50$ kW m$^{-2}$ at 1 mbar and 300 K. Therefore, we expect that light induced erosion occurs as long as $Ip/T > 15$ kW m$^{-2}$ Pa K$^{-1}$ \citep{wurm2007}. \\
In a minimum mass solar nebula \citep{hayashi1985} with given pressure, temperature and luminosity dependence this condition is reached at a distance of about 0.4 AU or Mercury's orbit \citep{wurm2007}.\\
For observed extrasolar systems we only know the luminosities for certain, not the pressure or temperature at the midplanes. To determine the distance of the dividing line between erosion and stability for the different objects, we assume that the dividing line is where the light flux of the host star equals that of a solar-type star in a minimum mass solar nebula or $0.4 $ AU $ \times \sqrt{L / L_{\text{sun}}}$ (Table \ref{tab:table1}). The implicit assumptions in this equation are that the gas pressure at this distance and the temperature are the same as in the minimum-mass solar nebula at 0.4 AU. \\
For the temperature this is a consistent assumption as the blackbody radiation temperature which defines the temperature of the disk in the minimum-mass solar nebula would be the same at the same incoming light flux. The pressure is less certain but one might argue that more luminous stars have denser disks. We therefore assume the pressure to be equal to that in a minimum mass solar nebula at the dividing line. However, more complex models would not change the following general statement.\\
We always find the erosion limit within the gap of the pre-transitional disks. In general, erosion is possible at the inner edges of full disks, although melting or sintering of particle ensembles might be an issue which has to be studied in further detail. For full disks and pre-transitional disks erosion might clearly provide the small dust observed over the lifetime of the disk. As the erosion limit distance is closer to the star than the outer wall in pre-transitional disks, particles released by the erosion mechanism can drift outward to this second wall by photophoresis and enhance the particle density locally \citep{krauss2005, haack2007}. This will induce an enhanced growth of dust aggregates. At the outer wall the growing bodies are safe from destruction unless they are scattered or drift inward again to the inner erosion zone during further evolution and recycling.

The holes of transitional disks are all larger than the erosion limit distance. As they have a largely reduced density it is unclear to what extent erosion will work here. The erosion limit distance would shift inward significantly. Earlier calculations by \citet{wurm2007} showed that, for example, in the case of TW Hya erosion close to the star would still be possible but this strongly depends on the gas content. The erosion zone likely decreases in thickness with time. However, even if it were only a very thin zone, dusty bodies traversing it might be significantly eroded. In the optical but thin still slightly gaseous hole, the dust would be transported by photophoresis and locally increase the particle density again. 

Therefore, at early full disk times erosion will influence the particle recycling at the inner edge. In later pre-transitional and transitional evolution times of the clearing phase the light induced erosion will be an important factor in shaping the size distribution or even triggering later formation of larger bodies at the outward moving edge.\\

The experiments reported here show that the number of ejected particles increases strongly with decreasing $g$-levels. The experiments performed on a parabolic flight display an inverse dependence on gravity and specific cohesion. The ejection rate is  higher by a factor two in 0.16$g$ than in 1$g$ for the basalt sample used. 
The importance of gravity deduced from these data can more clearly be seen in the model. If cohesion is reduced the ejection rate might be up to one or two orders of magnitude 
more efficient at low-$g$. Lower cohesion does not change the ejection rate strongly at high $g$-levels. 

As a consequence of microgravity on small planetesimals or sub-planetesimals, the effect is more intense than observed from ground-based experiments so far. \citet{kelling2011a} estimated from 1$g$ experiments a mass-loss rate of 10$^{-5}$ kg s$^{-1}$ m$^{-2}$ or about 10$^3$ kg year$^{-1}$ for an area 1 m$^2$ in size. Low gravity conditions would boost this value by a factor of a few. This compares well to the mass of a porous dusty body of meter size with a density of 1 g cm$^{-3}$. The total mass is some $10^{3}$ kg. It is therefore possible to erode, in less than one year, a significant part of such a body if it consists of loosely bound dust close to the disk edge. As the erosion zone is a fraction of an AU \citep{wurm2007} the experiments show that an object that passes the edge will efficiently lose mass (Figure \ref{fig:fig9}). Even for objects that lose mass but are not completely destroyed the effect is important as it provides a non-gravitational force on the body and will change the orbital parameters; the details, however, are beyond the scope of this paper. 

The mechanism is more efficient in an atmosphere consisting of hydrogen instead of air. The Equation \ref{eq:roha2} by \citet{rohatschek1995} for the photophoretic force shows a dependence on different parameters. The maximum force is proportional to a gas dependent parameter
\begin{equation}
D = \frac{\pi}{2} \sqrt{\frac{\pi}{3}\kappa} \frac{c \eta}{T}.
\end{equation}
 This parameter contains the dynamic viscosity of the gas $\eta$ and the average thermal velocity c, which is proportional to the square root of 1/$\mu$, the inverse molar mass. Assuming $\eta_{\text{air}}$ = 17.1 $\mu$Pa$\cdot$s and $\eta_{H_2}$ = 8.4 $\mu$Pa$\cdot$s, as well as $\mu_{\text{air}}$  = 29 g mol$^{-1}$ and $\mu_{\text{proto}}$ = 2.31 g/mol, we get $D_{\text{proto}}$  = 1.75 $\times$ $D_{\text{air}}$. The maximum force is therefore 1.75 times higher in a protoplanetary disk than on Earth.

In principle the erosion mechanism might compete with regrowth as a body moves through the disk and collects dust particles. This is an interesting aspect as it changes the surface morphology and the susceptibility for re-erosion. To what degree this process is important depends on the position of the dividing line with respect to the inner edge of the optically thick disk. If erosion only occurs in the transitional zone far away from the edge then this whole zone is optically thin by definition and re-growth is not important. This might be explained as follows. Assuming all dust particles to be spherical with a radius of 5 $\mu$m, the particle cross section is 78.5 $\times$ 10$^{-12}$ m$^2$. The disk would be optically thick if its cross section were completely covered with dust particles. At typical distances of 0.5 AU and with a scale height of 0.1 AU this would be 7 $\times$ 10$^{21}$ m$^2$. Dividing by the particle cross section, this gives 9 $\times$ 10$^{31}$ particles. Spread out over an estimated 0.5 AU in distance the particle density would be $n = 0.17$ particles m$^{-3}$. The largest drift velocities considered in disk models are on the order of 50 m s$^{-1}$. If a body of 1 m radius colleced dust at these speeds the growth rate would be 26.7 particle per second or with a particles density of 2000 kg m$^{-3}$ it would be 3 $\times$ 10$^{-11}$ kg s$^{-1}$. Compared to the erosion rate this worst case estimate shows that re-growth is negligible.\\
This changes at the edge. If the transition zone is only, for example, 100,000 km thick the particle density would be 127.3 particles m$^{-3}$ and the mass growth 2 $\times$ 10$^{-8}$ kg s$^{-1}$. This is a factor of 500 less than the erosion rate under perfect conditions, but for non-perfect conditions, e.g., further toward the dark side of the edge growth and erosion might be comparable. \\

Another application of the effect at low gravity is found on Mars. By adjusting the conditions (light flux and temperature) to Mars, dust entrainment in the atmosphere might be explained \citep{wurm2008}. The 
low-gravity experiments suggest higher erosion rates on Mars.\\ 

Most dust samples that absorb visible radiation show light-induced ejection. We have quantified the gravity effect for a single sample so far. We have experimental evidence that cohesion in the context of this ejection mechanism can vary strongly for different dust samples, but quantification is subject to future research. Overall, erosion of a dusty body at the low-gravity conditions of small bodies at the inner edges of protoplanetary disks is a major recycling process of dust, and is important in the context of planet formation.

\begin{acknowledgments}
This work was supported by the Deutsches Zentrum f\"ur Luft- und Raumfahrt (DLR), the Federal Ministry of Economics and Technology Germany (BMWi), the European Space Agency (ESA) and the Deutsche Forschungsgemeinschaft (DFG). \\
We thank the anonymous reviewer for valuable comments.
\end{acknowledgments}

\end{document}